\documentclass[a4paper,11pt]{article}
\pdfoutput=1 

\usepackage{jheppub} 
\usepackage{physics}
\usepackage{amsmath}
\usepackage{bbm}
\usepackage{xcolor}

\usepackage{tensor} 
\newcommand{\ind}[1]{\indices{#1}}

\newcommand\fft[2]{\frac{#1}{#2}}

\newcommand\nn{\nonumber}

\def\be{\begin{equation}}
\def\ee{\end{equation}}
\def\bse{\begin{subequations}}
\def\ese{\end{subequations}}

\def\zb{\bar{z}}

\def\calB{\mathcal{B}}
\def\calA{\mathcal{A}}
\def\calG{\mathcal{G}}

\def\bG{\overline{G}}
\def\bP{\overline{P}}
\def\bB{\overline{B}}
\def\bC{\overline{C}}

\usepackage[T1]{fontenc} 

\title{Gauged Six-Dimensional Supergravity from Warped IIB Reductions}

\author{Junho Hong,}
\author{James T. Liu,}
\author{Daniel R. Mayerson}

\affiliation{Leinweber Center for Theoretical Physics, Randall Laboratory of Physics \\The University of Michigan, Ann Arbor, MI 48109-1040, USA }

\emailAdd{junhoh@umich.edu}
\emailAdd{jimliu@umich.edu}
\emailAdd{drmayer@umich.edu}

\makeatletter
\newcommand*{\rom}[1]{\expandafter\@slowromancap\romannumeral #1@}
\makeatother

\abstract{We find a family of complete non-linear Kaluza-Klein reduction ans\"atze from type IIB supergravity to Romans' 6D $F(4)$ gauged supergravity in the bosonic sector. The reduction is over a sphere $S^2$ and a Riemann surface $\Sigma$, and depends on a pair of arbitrary locally holomorphic functions $\calA_{\pm}$ on $\Sigma$. This family of reductions is inspired by the recent construction of 1/2 BPS supersymmetric warped $AdS_6$ solutions of IIB supergravity that depend on these same functions $\calA_{\pm}$.}

\begin{document}
\maketitle
\flushbottom

\section{Introduction}

Supergravity solutions of the form $AdS_d\times S^n$ have long been known, and have particular importance in the context of AdS/CFT.  Perhaps the most familiar cases are those associated with the near-horizon geometry of M2, D3 and M5 branes.  In contrast, the case of $AdS_6/CFT_5$ has been less well explored, in part because it does not admit a single brane interpretation.  Nevertheless, there has been a recent resurgence of interest in this case, driven both on the field theory and the gravity sides of the duality.  In contrast with the more familiar cases, holography in this dimension is interesting as there are no maximally supersymmetric 5D SCFTs (with 16 Poincar\'e supercharges and 16 superconformal supercharges).  In particular, the maximal possible amount of supersymmetry in 5D is 8+8 supercharges \cite{Nahm:1977tg}. The superconformal algebra in 5D is then based on the Lie superalgebra $F(4)$ with maximal bosonic subalgebra $SO(2,5)\times SO(3)$ \cite{Minwalla:1997ka}.  The natural six-dimensional dual is then $F(4)$ gauged supergravity which is a non-chiral theory with 16 real supercharges \cite{Romans:1985tw}.

Of course, a string theory realization of $AdS_6/CFT_5$ starts not in six dimensions, but rather with 10 or 11 dimensional supergravity (in its low-energy limit) compactified to $AdS_6$ times some internal manifold.  Such backgrounds can preserve at most half of the total supersymmetries \cite{Figueroa-OFarrill:2002ecq}, in accordance with the $CFT_5$ picture.  Until recently, string theory realizations of such $AdS_6$ duals have been hard to come by, with the prime example being a construction of stacks of D4-branes, D8-branes, and O8-planes in type IIA string theory \cite{Seiberg:1996bd,Brandhuber:1999np,Bergman:2012kr,Passias:2012vp}. The presence of D8-branes means that the supergravity background is obtained in the massive IIA theory.

In type IIB, $(p,q)$ five-brane webs \cite{Aharony:1997ju,Aharony:1997bh} with D7-branes added \cite{DeWolfe:1999hj} can realize large classes of 5D SCFTs.  The holographic dual of these five-brane webs corresponds to supersymmetric $AdS_6$ solutions of IIB supergravity.  The Killing spinor equations for this system were investigated in \cite{Apruzzi:2014qva,Kim:2015hya} and reduced to a pair of coupled PDEs, and a complete family of local solutions was constructed in \cite{DHoker:2016ujz,DHoker:2017mds,DHoker:2017zwj}. The local $AdS_6$ solutions were found by studying the IIB Killing spinor equations, which led to a family of solutions of the form $AdS_6\times S^2$ warped over a Riemann surface $\Sigma$; remarkably, these solutions are completely determined by a pair of holomorphic functions $\calA_{\pm}$ on $\Sigma$ which can be chosen freely up to global regularity conditions.

More generally, the $AdS_6$ solutions of \cite{DHoker:2016ujz} ought to be viewed as vacuum solutions of six-dimensional $F(4)$ gauged supergravity obtained by reducing ten-dimensional IIB supergravity on $S^2\times\Sigma$.  Curiously, the existence of a whole family of solutions suggests that the lifting of $F(4)$ gauged supergravity to ten dimensions is far from unique.  This is in contrast to the 32 supercharge cases such as gauged $\mathcal N=8$ supergravity in five-dimensions, which has a unique lift to IIB supergravity on $S^5$.  Before this family of IIB vacua was discovered, a full non-linear Kaluza-Klein reduction of massive IIA supergravity to $F(4)$ gauged supergravity on (one hemisphere of) $S^4$ was obtained in \cite{Cvetic:1999un}.  This was subsequently dualized to a IIB reduction in \cite{Jeong:2013jfc} using non-abelian T-duality.  In particular, the IIA reduction ansatz of \cite{Cvetic:1999un} was based on $AdS_6$ times a squashed $S^4$ foliated by 3-spheres, with the $SU(2)$ $R$-symmetry corresponding to the gauging of $SU(2)_L$ inside the $SO(4)\simeq SU(2)_L\times SU(2)_R$ isometry of $S^3$.  Non-abelian T-dualizing then gives rise to a IIB reduction where the $S^4$ is replaced by $S^2\times\Sigma$, with the $R$-symmetry now corresponding to the $SO(3)$ isometry of $S^2$.  However, this procedure gives rise to only a single background, first obtained in \cite{Lozano:2012au}, and not the entire family of solutions constructed in \cite{DHoker:2016ujz}.

In this paper, we present a complete family of consistent truncations of IIB supergravity to $F(4)$ gauged supergravity parametrized by the same holomorphic functions $\calA_\pm$ of \cite{DHoker:2016ujz}. This generalizes the family of $AdS_6$ vacua into complete reduction ans\"atze, in agreement with the conjecture that any supersymmetric vacuum solution can be extended to a consistent truncation incorporating the full lower-dimensional supergravity multiplet \cite{Duff:1985jd,Gauntlett:2007ma}.  (General half-maximal consistent truncations were recently considered in the framework of exceptional field theory in \cite{Malek:2017njj}.)  The non-linear reduction to $F(4)$ gauged supergravity was inspired by the construction of \cite{Jeong:2013jfc}, but is in fact much more general.  Naturally, it incorporates the reduction ansatz of \cite{Jeong:2013jfc} as a special case, and we give the explicit functions $\calA_{\pm}$ necessary to recover the results of \cite{Jeong:2013jfc} in section \ref{sec:discussion}. 

The rest of the paper is structured as follows. In section \ref{sec:10Dsetup}, we review the basics of 10D type IIB supergravity, and briefly discuss the 1/2 BPS $AdS_6\times S^2\times \Sigma$ solutions of \cite{DHoker:2016ujz} that depend on the pair of holomorphic functions $\calA_{\pm}$ on $\Sigma$. Then, in section \ref{sec:6Dsugra+reduction}, we discuss Romans' 6D $F(4)$ gauged supergravity and give our full non-linear reduction ansatz, which also depends on the holomorphic functions $\calA_{\pm}$. Finally, in section \ref{sec:discussion}, we discuss how our ansatz reduces to previous results in the literature in special cases, and the connection to $AdS_6/CFT_5$ holography. The appendices discuss our conventions for indices and coordinates on the various manifolds we consider (appendix \ref{app:ourconventions}), the relation between different conventions for IIB supergravity (appendix \ref{app:10Dconventions}), and more details of checking the equations of motion on our ansatz (appendix \ref{app:EOMchecks}).

\paragraph{Main result:} Our main result is the full non-linear Kaluza-Klein reduction ansatz given in section \ref{sec:genansatz} in (\ref{general:ansatz:1})-(\ref{general:ansatz:G5}). On this ansatz, the 10D type IIB supergravity equations of motion (\ref{2B:eoms}) are equivalent to the 6D $F(4)$ gauged supergravity equations of motion (\ref{F(4):eoms}). The reduction ansatz depends on a pair of unrestricted holomorphic functions $\calA_{\pm}$ on the Riemann surface $\Sigma$.

\paragraph{Note added:} After this work appeared on the arXiv, the work \cite{Malek:2018zcz} was released, which also constructs the consistent truncation to $F(4)$ gauged supergravity.

\section{IIB supergravity and warped \texorpdfstring{$AdS_6\times S^2\times\Sigma$}{AdS6 x S2 x Sigma} vacua} \label{sec:10Dsetup}

Before presenting the non-linear Kaluza-Klein ansatz, we quickly review the equations of motion of IIB supergravity and the family of supersymmetric $AdS_6$ solutions of \cite{DHoker:2016ujz}.  This sets up our conventions and lays the groundwork for the reduction.

\subsection{Type IIB supergravity}

Type IIB supergravity is a chiral theory with 32 real supercharges.  Because of the self-dual five-form, it does not admit a covariant Lagrangian formulation (although one can come close).  Of course, for presenting the consistent truncation, we only need the equations of motion, which we give in the $SU(1,1)$ formulation \cite{Schwarz:1983qr,Howe:1983sra}.

The bosonic fields consist of the Einstein frame metric $g_{MN}$, a complex scalar with kinetic term $P_1$ and composite connection $Q_1$, a complex three-form field strength $G_3$ and a self-dual five-form $F_5=*F_5$.  From a stringy point of view, the one-forms $P$ and $Q$ are related to the dilaton and RR axion, while $G_3$ contains both NSNS and RR (real) three-forms.  The exact map is spelled out in appendix \ref{app:10Dconventions}.

The IIB supergravity fields satisfy the Bianchi identities:
\bse
\begin{align}
	dP&=2iQ\wedge P,\\
	dQ&=-iP\wedge\bP,\\
	dG_3&=iQ\wedge G_3-P\wedge\bG_3.
\end{align}
\ese
These are automatically satisfied if we introduce a complex scalar $B$ and a complex two form $C_2$ as:
\bse
	\begin{align}
\label{eq:10D:PforB}	P&=(1-|B|^2)^{-1}dB,\\
\label{eq:10D:QforB}	Q&=(1-|B|^2)^{-1}\Im[Bd\bB],\\
\label{eq:10D:G3forC2}	G_3&=(1-|B|^2)^{-\frac{1}{2}}(dC_2-Bd\bC_2).
	\end{align}
\label{eq:10D:PG3def}\ese
The equations of motion are:
\bse
\begin{align}
	(d-2iQ)\wedge*P&=-\frac{1}{4}G_3\wedge* G_3,\label{2B:eom:P}\\
	(d-iQ)\wedge*G_3&=P\wedge*\bG_3-4iG_3\wedge* F_5,\label{2B:eom:G3}\\
    d*F_5&=\fft{i}8G_3\wedge\bG_3,\label{2B:eom:F5}
\end{align}
    \label{2B:eoms}
\ese
along with the Einstein equation
\begin{align}
	R_{IJ}&=P_I\bP_J+\bP_I P_J+\frac{1}{6}F_{IKLMN}F\ind{_J^{KLMN}}\nn\\
	&\quad+\frac{1}{8}\left(G_{IKL}\bG\ind{_J^{KL}}+\bG_{IKL}G\ind{_J^{KL}}\right)-\frac{1}{48}g_{IJ}G_{KLM}\bG^{KLM}.\label{2B:eom:Einstein}
\end{align}
Note that the five-form equation can equally well be viewed as a Bianchi identity, $dF_5=\fft{i}8G_3\wedge\bG_3$, which may be solved by taking $F_5=dC_4+\fft{i}8\Im[C_2\wedge d\bC_2]$.  However, as in other IIB reductions, it is more convenient to make the ansatz on the field strength, and not the potential, in which case the equation of motion (\ref{2B:eom:F5}) will need to be verified.

\subsection{A family of warped \texorpdfstring{$AdS_6$}{AdS6} solutions}\label{sec:SUSYAdS6sols}

A large family of 1/2 BPS solutions to IIB supergravity of the form $AdS_6\times S^2$ warped over a 2D Riemann surface $\Sigma$ (with complex coordinates $z,\zb$) were found by a detailed analysis of the IIB supersymmetry equations in \cite{DHoker:2016ujz}. These solutions are (locally) completely determined by a pair of holomorphic functions $\calA_{\pm}(z)$ on the Riemann surface. Remarkably, there is no restriction on the pair of holomorphic functions in order to obtain a local supersymmetric solution.\footnote{A different but presumably equivalent classification of these $AdS_6$ solutions was found in \cite{Apruzzi:2018cvq}. These solutions were given in terms of solutions to the spherically symmetric cylindrical Laplace equation instead of holomorphic functions.} To obtain a well-behaved global solution, considerably more care is needed: $\calA_{\pm}$ are allowed to be multi-valued as holomorphic sections of a holomorphic bundle over $\Sigma$ with structure group contained in $SU(1,1)\times \mathbb{C}$; the boundaries and other global properties of $\Sigma$ are important in constructing such solutions \cite{DHoker:2016ujz,DHoker:2017mds,DHoker:2017zwj,DHoker:2016ysh}. There has been much interest in these global solutions and interpreting them as near-horizon limits of $(p,q)$ 5-brane webs with additional 7-branes in the web \cite{DHoker:2017mds,DHoker:2017zwj,DHoker:2016ysh,Gutperle:2018vdd,Bergman:2018hin}. (See also section \ref{sec:discussionholo} for more discussion regarding the holographic duals of these solutions.)

Here we briefly summarize the solutions of \cite{DHoker:2016ujz}.  The starting point is a pair of holomorphic functions $\calA_\pm(z)$, from which we may define a holomorphic function $\calB$ whose derivative is given by (using obvious notation $\partial=\partial_z)$:
\be \label{eq:Bdef} \partial_z \calB = \calA_+\partial\calA_- - \calA_- \partial \calA_+.\ee
Of the undetermined integration constant in $\calB$, only the real part is relevant.  From $\calA_\pm$ and $\calB$, we define
\begin{equation}
\kappa_\pm=\partial\mathcal A_\pm,\qquad\kappa^2=-|\kappa_+|^2+|\kappa_-|^2,
\end{equation}
along with
\begin{equation}
\mathcal G=|\mathcal A_+|^2-|\mathcal A_-|^2+\mathcal B+\bar{\mathcal B}.
\end{equation}
Note that $\kappa^2 = -\partial\bar{\partial}G$.  We also find it useful to introduce
\begin{equation}
    \mathcal{Y}= \fft{\kappa^2\mathcal G}{|\partial\mathcal G|^2},\qquad
	\hat{\mathcal C}=\fft{\kappa_+\bar\partial\mathcal G+\bar\kappa_-\partial\mathcal G}{\kappa^2}.
	\label{eq:YCdef}
\end{equation}

The supersymmetric $AdS_6$ vacua of \cite{DHoker:2016ujz} are then given by the metric
\begin{equation}
    ds^2 = f_6^2 ds^2_{AdS_6} +f_2^2 ds^2_{S^2} + 4\rho^2 dz d\zb,
\label{eq:susysolmet}
\end{equation}
with metric functions
\begin{equation}
     f_6^2=\fft{c_6^2}{\rho^2}\kappa^2\sqrt{\tilde{\mathcal D}},\qquad
     f_2^2=\fft{c_6^2}{9\rho^2}\fft{\kappa^2}{\sqrt{\tilde{\mathcal  D}}},\qquad
    \rho^4=\fft{c_6^2}6\fft{\kappa^4\sqrt{\tilde{\mathcal  D}}}{\mathcal G},
\label{eq:metfuncs}
\end{equation}
and matter fields
\begin{subequations}
\begin{align}
 B &= \fft{(\mathcal A_+-\bar{\mathcal A}_-)-\hat{\mathcal C}/\sqrt{\tilde{\mathcal  D}}}{(\bar{\mathcal A}_+-\mathcal A_-)+\bar{\hat{\mathcal C}}/\sqrt{\tilde{\mathcal D}}}, \label{eq:Bvac}\\
 C_2 &= \frac{2ic_6}{9}\left[\fft{\hat{\mathcal C}}{\tilde{\mathcal D}}-3(\bar{\mathcal A}_-+\mathcal A_+)\right]\mathrm{vol}(S^2),\label{eq:Cvac}\\
F_5 &= 0,\label{eq:Fvac}
\end{align}%
\label{eq:susysolmat}%
\end{subequations}%
Here $c_6$ is a non-vanishing constant, and we have furthermore introduced the recurring factor
\begin{equation}
	\tilde{\mathcal  D}=1+\fft{2}{3\mathcal Y},
\label{eq:calD}
\end{equation}
which will play an important role in the generalization to the full KK reduction below.

Finally, note that there are in fact two branches of solutions, the first with
\be \label{eq:conditionskappaG} \kappa^2\geq 0, \qquad \mathcal{G}\geq 0,\ee
and the second with both quantities non-positive.  (These two branches of solutions are mapped into each other by complex conjugation \cite{DHoker:2016ujz}.)  Of course, the metric functions $f_2^2$, $f_6^2$ and $\rho^2$ must all be positive.  This will be the case provided we take the positive square-root ($\sqrt{\tilde{\mathcal D}}>0$) on the first branch and the negative square-root ($\sqrt{\tilde{\mathcal D}}<0$) on the second branch.

\section{6D Supergravity and the reduction ansatz}\label{sec:6Dsugra+reduction}

The existence of a family of $AdS_6$ solutions suggests the possibility of a complete non-linear Kaluza-Klein reduction of IIB supergravity to six-dimensional $F(4)$ gauged supergravity.  Here we first present a brief overview of the six-dimensional theory, and then turn to the full reduction ansatz generalizing the vacuum solution discussed above in section \ref{sec:SUSYAdS6sols}.

\subsection{\texorpdfstring{$F(4)$}{F(4)} gauged supergravity}\label{sec:6Dsugra}

Romans' six-dimensional $F(4)$ gauged supergravity \cite{Romans:1985tw} is a non-chiral theory with 16 real supercharges.  The bosonic fields consist of a metric $\tilde g_{\mu\nu}$, a real scalar $\tilde{\phi}$, an Abelian two-form $\tilde F_2$ and three-form $\tilde F_3$, and three $SU(2)$ gauge two-forms $\tilde F^i$. The bosonic Lagrangian may be written in a form notation \cite{Romans:1985tw,Cvetic:1999un} as
\begin{align}
    \mathcal L&=R*_6\mathbbm1-4\fft{*_6dX\wedge dX}{X^2}-\tilde g^2\left(\fft29X^{-6}-\fft83X^{-2}-2X^2\right)*_61\nn\\
    &\qquad-\fft12X^4*_6\tilde F_3\wedge\tilde F_3-\fft12X^{-2}\left(*_6\tilde F_2\wedge\tilde F_2+\fft1{\tilde g^2}*_6\tilde F^i\wedge\tilde F^i\right)\nn\\
    &\qquad-\tilde A_2\wedge\left(\fft12d\tilde A_1\wedge d\tilde A_1+\fft13\tilde g\tilde A_2\wedge d\tilde A_1+\fft2{27}\tilde g^2\tilde A_2\wedge\tilde A_2+\fft1{2\tilde g^2}\tilde F^i\wedge\tilde F^i\right),
    \label{eq:F4lag}
\end{align}
where $X=e^{-\frac{1}{2\sqrt{2}}\tilde\phi}$, and the field strengths are given in terms of the gauge potentials by
\bse
\begin{align}
	\tilde F_3&=d\tilde A_2,\\
	\tilde F_2&=d\tilde A_1+\frac{2}{3}\tilde g\tilde A_2,\\
	\tilde F^i&=d\tilde A^i+\frac{1}{2}\epsilon_{ijk}\tilde A^j\wedge\tilde A^k.
\end{align}
\ese
The equations of motion corresponding to this Lagrangian are given by
\bse
	\begin{align}
	d(X^4*_6\tilde F_3)&=-\frac{1}{2}\tilde F_2\wedge\tilde F_2-\frac{1}{2\tilde g^2}\tilde F^i\wedge\tilde F^i-\frac{2}{3}\tilde gX^{-2}*_6\tilde F_2,\label{F(4):eom:1}\\
	d(X^{-2}*_6\tilde F_2)&=-\tilde F_2\wedge\tilde F_3,\label{F(4):eom:2}\\
	D(X^{-2}*_6\tilde F^i)&=-\tilde F_3\wedge\tilde F^i,\label{F(4):eom:3}\\
	d(X^{-1}*_6dX)&=\frac{1}{8}X^{-2}\left(*_6\tilde F_2\wedge\tilde F_2+\frac{1}{\tilde g^2}*_6\tilde F^i\wedge\tilde F^i\right)-\frac{1}{4}X^4*_6\tilde F_3\wedge\tilde F_3\nonumber\\&\quad+\tilde g^2\left(\frac{1}{6}X^{-6}-\frac{2}{3}X^{-2}+\frac{1}{2}X^2\right)*_6\mathbbm{1},\label{F(4):eom:4}
	\end{align}
\label{F(4):eoms}\ese
along with the Einstein equation
\begin{align}
	\tilde R_{\mu\nu}&=4X^{-2}\partial_\mu X\partial_\nu X+\tilde g^2\left(\frac{1}{18}X^{-6}-\frac{2}{3}X^{-2}-\frac{1}{2}X^2\right)g_{\mu\nu}+\frac{1}{4}X^4\left((\tilde F_3)^2_{\mu\nu}-\frac{1}{6}g_{\mu\nu}(\tilde F_3)^2\right)\nonumber\\&\quad+\frac{1}{2}X^{-2}\left((\tilde F_2)^2_{\mu\nu}-\frac{1}{8}g_{\mu\nu}(\tilde F_2)^2\right)+\frac{1}{2\tilde g^2}X^{-2}\left((\tilde F^i)^2_{\mu\nu}-\frac{1}{8}g_{\mu\nu}(\tilde F^i)^2\right).\label{F(4):eom:5}
\end{align}
In the above, the $SU(2)$ gauge covariant derivative $D$ is defined by
\begin{equation}
	D\tilde F^i=d\tilde F^i+\epsilon_{ijk}\tilde A^j\wedge\tilde F^k.
\end{equation}

This $F(4)$ gauged supergravity theory admits a supersymmetric $AdS_6$ vacuum with $\tilde F_2=\tilde F_3=\tilde F^i = 0$ and $X=1$ that preserves all 16 of the supersymmetries \cite{Romans:1985tw}. There is also a non-supersymmetric $AdS_6$ vacuum with $X = 3^{-1/4}$. Note that to get a unit radius $AdS_6$, we must choose:
\be \tilde{g} = \frac{3}{\sqrt{2}}.
\label{eq:tildeg}
\ee
We will make this choice from here on.

\subsection{Generalized reduction ansatz}\label{sec:genansatz}

The $F(4)$ gauged supergravity theory was obtained from a warped $S^4$ reduction of massive IIA supergravity in \cite{Cvetic:1999un}, and more recently from a reduction of IIB supergravity over an $S^2$ and a Riemann surface in \cite{Jeong:2013jfc}.  In both cases, the supersymmetric $AdS_6$ vacua uplifted to 10D are 1/2 BPS solutions of either IIA or IIB supergravity.

We now present our main result, which is a consistent truncation of type IIB supergravity to 6D $F(4)$ gauged supergravity generalizing the reduction of \cite{Cvetic:1999un,Jeong:2013jfc}.  We start with the metric
\begin{equation}
    ds^2=f_6^2ds_6^2+f_2^2ds^2_{\tilde S^2} +4\rho^2dzd\zb,\label{general:ansatz:1}
\end{equation}
where
\begin{equation}
    f_6^2=\fft{c_6^2}{\rho^2}\kappa^2\sqrt{\mathcal D},\qquad
    f_2^2=\fft{c_6^2}{9\rho^2}\fft{\kappa^2X^2}{\sqrt{\mathcal D}},\qquad
    \rho^4=\fft{c_6^2}{6}\fft{\kappa^4X^2\sqrt{\mathcal D}}{\mathcal G}.
    \label{eq:fXs}
\end{equation}
The matter fields are given by
\begin{subequations}
\begin{align}
	B&=\fft{(\mathcal A_+-\bar{\mathcal A}_-)-\hat{\mathcal C}X^2/\sqrt{\mathcal D}}{(\bar{\mathcal A}_+-\mathcal A_-)+\bar{\hat{\mathcal C}}X^2/\sqrt{\mathcal D}},\label{general:ansatz:2}
	\\
	C_2&=\frac{2ic_6}{9}\left[\fft{\hat{\mathcal C}}{\mathcal D}-3(\bar{\mathcal A}_-+\mathcal A_+)\right]\mathrm{vol}(\tilde S^2)-\sqrt2c_6(\bar{\mathcal A}_--\mathcal A_+)\tilde F_2+\fft{2ic_6}3(\bar{\mathcal A}_-+\mathcal A_+)\mu^i\tilde F^i,\label{general:ansatz:3}\\
	F_5&=G_5+*G_5,\label{general:ansatz:4}
\end{align}
\end{subequations}
with
\be \label{general:ansatz:G5} G_5=c_6^2\left[\fft{i\kappa^2X^4}{2}(*_6\tilde F_3)\wedge dz\wedge d\bar z+\fft{1}{2\sqrt{2}X^2}(*_6\tilde F_2)\wedge*_2d\mathcal G-\fft{1}{6X^2}(*_6\tilde F^i)\wedge D(\mu^i\mathcal G)\right].\ee
It is also useful to give the expression for $*G_5$:
\begin{align}
	*G_5&=c_6^2\Bigg[\bigg(\fft{\mathcal GX^4}{6\mathcal D}\tilde F_3+\fft{\sqrt{2}}{36\mathcal D}\tilde F_2\wedge d\mathcal G+\fft{1}{54\mathcal D}\mu^i\tilde F^i\wedge *_2d\mathcal G\bigg)\wedge\mathrm{vol}(\tilde S^2)\nn\\&\kern3em+\fft{i\kappa^2}{18}\tilde F^i\wedge*_2D\mu^i\wedge dz\wedge d\zb\Bigg].
\end{align}

Here $\mathcal{D}$ is defined as
\be
    \mathcal D=X^4+\fft{2}{3\mathcal Y},
\label{eq:newcalD}
\ee
which generalizes (\ref{eq:calD}) in the presence of a non-trivial scalar. Note that $ds^2_{\tilde S^2}$ and $\mathrm{vol}(\tilde S^2)$ have been used instead of $ds^2_{S^2}$ and $\mathrm{vol}(S^2)$ since the $SU(2)$ isometry of the $S^2$ is gauged by the $\tilde A^i$ fields. (See appendix \ref{app:ourconventions} for more information.) The remaining definitions, (\ref{eq:Bdef}) through (\ref{eq:YCdef}), are unchanged. 

Our reduction ansatz thus provides a consistent truncation of IIB supergravity on an $S^2$ and warped over a Riemann surface $\Sigma$ to 6D $F(4)$ gauged supergravity. In fact, our ansatz contains a family of such truncations --- one for each pair of holomorphic functions $\calA_{\pm}$ on the Riemann surface. This fully generalizes and contains the supersymmetric $AdS_6$ solutions of \cite{DHoker:2016ujz} discussed in section \ref{sec:SUSYAdS6sols}.  In particular, (\ref{eq:fXs}) and (\ref{general:ansatz:2}) generalize the corresponding expressions (\ref{eq:metfuncs}) and (\ref{eq:Bvac}) to incorporate the scalar $X$, and (\ref{general:ansatz:3}) and (\ref{general:ansatz:4}) generalize (\ref{eq:Cvac}) and (\ref{eq:Fvac}) to incorporate the six-dimensional gauge fields.  When $X=1$ (i.e.\ the scalar $\tilde{\phi}=0$) and the gauge fields are turned off, $\tilde F_2 = \tilde F_3 = \tilde F^i = 0$, this reduction ansatz reduces as it must to that of \cite{DHoker:2016ujz}.

We have explicitly checked that our ansatz (\ref{general:ansatz:1})-(\ref{general:ansatz:G5}) satisfies the 10D IIB equations of motion (\ref{2B:eoms}) if and only if the 6D fields $\tilde g_{\mu\nu}$, $\tilde{F}_2$, $\tilde{F}_3$, $\tilde{F}^i$ and $X$ satisfy the 6D equations of motion (\ref{F(4):eoms}) and (\ref{F(4):eom:5}). In particular, the IIB axi-dilaton equation of motion (\ref{2B:eom:P}) is equivalent to the 6D scalar equation of motion (\ref{F(4):eom:4}), the IIB three-form equation (\ref{2B:eom:G3}) is equivalent to (\ref{F(4):eom:1}), (\ref{F(4):eom:2}), (\ref{F(4):eom:3}), and (\ref{F(4):eom:4}), and the self-dual five-form equation (\ref{2B:eom:F5}) is equivalent to (\ref{F(4):eom:1}), (\ref{F(4):eom:2}), and (\ref{F(4):eom:3}). Finally, the IIB Einstein equation, (\ref{2B:eom:Einstein}), is satisfied if and only if (\ref{F(4):eom:3}), (\ref{F(4):eom:4}) and the 6D Einstein equation (\ref{F(4):eom:5}) are satisfied. For more details of these calculations, see appendix \ref{app:EOMchecks}.

Our ansatz thus gives a family of consistent truncations of 10D IIB supergravity to 6D $F(4)$ gauged supergravity, in the sense that any solution of the 6D theory can be uplifted to a family of 10D solutions using (\ref{general:ansatz:1})-(\ref{general:ansatz:G5}). We stress once more that this ansatz is consistent for \emph{any} pair of holomorphic functions $\calA_{\pm}$, at least up to regularity conditions of the $AdS_6$ vacuum.

\section{Discussion}\label{sec:discussion}

The reduction ansatz (\ref{general:ansatz:1})-(\ref{general:ansatz:G5}) generalizes the IIB reduction of \cite{Jeong:2013jfc}, corresponding to the $AdS_6$ background of \cite{Lozano:2012au}, to encompass the complete family of $AdS_6$ backgrounds of \cite{DHoker:2016ujz}.  In fact, our method for finding this generalized ansatz was to translate the specific reduction of \cite{Jeong:2013jfc} into the language of the holomorphic functions $\calA_\pm$ of \cite{DHoker:2016ujz}.  For example, starting with the string frame metric embedding of \cite{Jeong:2013jfc}
\begin{equation}
	d\hat s^2=X^{-\frac{1}{2}}s^{-\frac{1}{3}}\Delta^{\frac{1}{2}}\left[ds_6^2+2\tilde g^{-2}X^2d\xi^2\right]+e^{-2A}dr^2+\frac{r^2e^{2A}}{r^2+e^{4A}}ds^2_{\tilde S^2},
	\label{particular:ansatz:met}
\end{equation}
with
\begin{equation}
	e^A=\frac{1}{\sqrt{2}\tilde g}s^{-\frac{1}{6}}c\,X^{-\fft34}\Delta^{-\fft14}, \qquad \Delta = X c^2 + X^{-3}s^2, \qquad s = \sin \xi, \quad c = \cos\xi,
\end{equation}
we deduce the coordinate transformation on the Riemann surface $\Sigma$ mapping $\{\xi,r\}\to\{z,\bar z\}$
\begin{equation}
    z=z_1+iz_2, \qquad z_1 = \frac23 \tilde{g}^2 r,\qquad z_2 = \sin^{2/3}\xi.
    \label{eq:particular:ansatz:coords}
\end{equation}
This allows us to rewrite (\ref{particular:ansatz:met}) as
\begin{equation}
    d\hat s^2=X^{-\frac{1}{2}}s^{-\frac{1}{3}}\Delta^{\frac{1}{2}}ds_6^2+\frac{r^2e^{2A}}{r^2+e^{4A}}ds^2_{\tilde S^2}+\fft9{4\tilde g^4}e^{-2A}dz\,d\bar z.
\end{equation}
Converting to the Einstein frame and comparing with the $AdS_6$ metric (\ref{general:ansatz:1}) then yields the $X$-dependent metric functions (\ref{eq:fXs}), provided we identify
\begin{equation}
    \kappa^2=\fft{z_1z_2}{36c_6^2},\qquad\calG=\fft{(1-z_2^3)z_1}{54c_6^2},\qquad|\partial\calG|^2=\fft{z_1^2z_2^4+\fft19(1-z_2^3)^2}{(36c_6^2)^2},
    \label{eq:kapcalG}
\end{equation}
where we have additionally set $\tilde g=3/\sqrt2$ as in (\ref{eq:tildeg}).

Working backwards from (\ref{eq:kapcalG}), it is not too difficult to deduce the form of the holomorphic functions $\calA_\pm$
\begin{align}\label{eq:particular:ansatz:calA}
	\mathcal A_\pm(z)&=\fft1{c_6}\left(\frac{1}{216}z^3\mp\frac{i}{4}z-\frac{i}{108}\right),
\end{align}
along with the auxiliary function
\begin{align}
	\mathcal B(z)&=\fft1{c_6^2}\left(-\frac{i}{864}z^4+\frac{1}{216}z\right).
\end{align}
Given these functions, it is then possible to work out (\ref{general:ansatz:2}) as the appropriate generalization of the IIB axi-dilaton reduction (\ref{eq:Bvac}) in the presence of a non-trivial scalar $X$.  The remaining expressions are then for the complex three-form $G_3$ and self-dual five-form $F_5$.  These take a bit more effort, but can be obtained by translating the particular ansatz of \cite{Jeong:2013jfc} into the $SU(1,1)$ Einstein frame and reexpressing the result in terms of the holomorphic functions $\calA_{\pm}$ and their derived quantities such as $\calG$ and $\kappa_{\pm}$. Finally, as it was not guaranteed that this construction would be a consistent truncation, it was essential to check the IIB equations of motion and verify that they were equivalent to the $F(4)$ gauged supergravity equations of motion --- and in particular, that they did not impose any extra conditions on the functions $\calA_{\pm}$.  Additional details of these checks are presented in appendix~\ref{app:EOMchecks}.

\subsection{Holography}\label{sec:discussionholo}

Finally, we conclude with a few remarks about $AdS_6/CFT_5$ holography.  From a 10D point of view, we may use the supersymmetric $AdS_6$ solutions of \cite{DHoker:2016ujz,DHoker:2017mds,DHoker:2017zwj} discussed in section \ref{sec:SUSYAdS6sols}, see e.g. \cite{DHoker:2016ysh, Gutperle:2017tjo, Gutperle:2018wuk,Gutperle:2018vdd, Bergman:2018hin, Fluder:2018chf}. These efforts have been limited to the supersymmetric $AdS_6$ vacua as the extension of these solutions to include non-BPS excitations were not known until now.

Alternatively, one can approach 6D/5D holography in supergravity from the 6D perspective; efforts using the 6D $F(4)$ supergravity discussed in section \ref{sec:6Dsugra} include \cite{Ferrara:1998gv,DAuria:2000afl,Karndumri:2012vh,Karndumri:2014lba,Alday:2014rxa,Alday:2014bta,Alday:2014fsa,Hama:2014iea,Alday:2015jsa,Gutperle:2017nwo,Chang:2017mxc,Gutperle:2018axv}. The 6D supergravity is then the effective theory that describes (a consistent truncation of the set of) excitations around the $AdS_6$ vacuum. However, without an understanding of the uplift of these 6D solutions to 10D, a microscopic understanding of the CFT described by the $AdS_6$ vacuum and its excitations was lacking.

Our reduction ansatz (\ref{general:ansatz:1})-(\ref{general:ansatz:G5}) provides a key link between the 10D and 6D approaches above. In the 10D approach, it provides a way to include (non-BPS) excitations to the family of $AdS_6$ vacua. From the 6D perspective, it gives a way to understand the microscopic dual 5D CFT through the brane picture obtained by the 10D uplift of the 6D solutions. For example, the fact that our uplift is independent of the choice of holomorphic functions $\calA_{\pm}$ shows that it is perhaps not particularly surprising that the massive spin-2 excitations around the supersymmetric $AdS_6$ vacua found in \cite{Gutperle:2018wuk} are universally present for any $\calA_{\pm}$.

It would be interesting to extend the ansatz (\ref{general:ansatz:1})-(\ref{general:ansatz:G5}) further to include more fields in the truncation from 10D. For example, one could consider coupling vector multiplets to the 6D $F(4)$ theory; such an ansatz would then provide a 10D uplift of the 6D Janus solutions found in \cite{Gutperle:2017nwo}, which should correspond to mass deformations of the dual 5D SCFTs.  It is not obvious, however, whether the consistent truncation obtained here can be generalized to include the addition of matter multiplets, as non-linear Kaluza-Klein consistency often requires a delicate balance between internal harmonics unless a symmetry argument can be made \cite{Duff:1984hn,Hoxha:2000jf}, which does not appear to be the case for the Janus solutions.  A possibly more fruitful approach would be to consider instead gauge fields that arise from D3-branes wrapping 3-cycles \cite{Bergman:2018hin}, as they are not subject to the standard constraints of Kaluza-Klein consistency.

\acknowledgments
This project was initiated following discussions with C.F.~Uhlemann at the 2018 USU Workshop on Strings and Black Holes. We wish to thank C.F.~Uhlemann for enlightening discussions and E.~O~Colgain and O.~Varela for useful comments. This work was supported in part by the U.S.~Department of Energy under grant DE-SC0007859.

\appendix

\section{Conventions}\label{app:ourconventions}
For 10D indices, we use $K, L, M, \ldots$, and for 6D indices we use $\mu, \nu, \rho, \ldots$.  We work in both six and ten dimensions with a mostly plus signature.  In particular, the $F(4)$ gauged supergravity Lagrangian given in (\ref{eq:F4lag}) has been converted from the original mostly minus signature of \cite{Romans:1985tw}.

The complex coordinates on the 2D Riemann surface $\Sigma$ are $z,\zb$. With respect to these coordinates, we define:
\be
*_2 dz = idz.
\label{eq:Sigma*}
\ee

On the two-sphere $S^2$, we can use (three) constrained coordinates $\mu^i$ which satisfy $\sum_{i=1}^3\mu^i\mu^i = 1$. The metric and the volume form on the $S^2$ are then: 
\begin{equation}
	ds^2_{S^2} =\sum_{i=1}^3 d\mu^i d\mu^i,\qquad\qquad\mathrm{vol}(S^2)=\fft12\epsilon_{ijk}\mu^id\mu^j\wedge d\mu^k,
\end{equation}
respectively. The $SU(2)$ gauge-covariant derivative $D$ is defined by:
\begin{equation}
	D\tilde F^i=d\tilde F^i+\epsilon_{ijk}\tilde A^j\wedge\tilde F^k.
\end{equation}
The natural metric and the volume form on the $S^2$ when the $SU(2)$ isometries are gauged by the $\tilde F^i$ gauge fields are then:
\begin{equation}
	ds^2_{\tilde S^2}=\sum_{i=1}^3 D\mu^i D\mu^i,\qquad\qquad\mathrm{vol}(\tilde S^2)=\fft12\epsilon_{ijk}\mu^iD\mu^j\wedge D\mu^k,
\end{equation}
respectively. We also define:
\be
*_2 D\mu^i = \epsilon_{ijk} \mu^j D\mu^k.
\label{eq:S2*}
\ee
Even though the same symbol $*_2$ is used for Hodge duality on the $S^2$ and $\Sigma$, it should be clear which is meant by the context it is used in. Note that an explicit form of the $\mu^i$ coordinates could be:
\bse
\begin{align}
 \mu^1 &= \sin\theta\sin\phi, & \mu^2 &= \sin\theta\cos\phi, & \mu^3 & =-\cos\theta,
\end{align}
\ese
so that the ungauged metric is the usual $ds^2_{S^2}= d\theta^2 + \sin^2\theta d\phi^2$.

Our conventions for Hodge duality are given by:
\begin{equation}
    *(dx^{\mu_1}\wedge\cdots\wedge dx^{\mu_r})=\frac{\sqrt{|g|}}{(D-r)!}\epsilon_{\nu_{1}\cdots \nu_{D-r}}{}^{\mu_1\cdots \mu_r}dx^{\nu_{1}}\wedge\cdots\wedge dx^{\nu_{D-r}},\label{Hodge:dual}
\end{equation}
where $D$ is the dimension.  The relation between 10D and 6D Hodge duality is given by
\begin{equation}
    \epsilon_{\mu_1\cdots \mu_6\alpha\beta ij}^{(10D)}=\epsilon_{\mu_1\cdots \mu_6}^{(6D)}\epsilon_{\alpha\beta}\epsilon_{ij}=1,
\end{equation}
where $\alpha$, $\beta$ are orthonormal coordinates for the $\tilde S_2$ and $i$, $j$ are real coordinates on $\Sigma$.  The orientation on $\tilde S_2$ and $\Sigma$ are given by (\ref{eq:S2*}) and (\ref{eq:Sigma*}), respectively.

\section{Mapping between different 10D IIB conventions}\label{app:10Dconventions}

Type IIB supergravity in 10D is often formulated in terms of the (real) axion $\hat{C}_0$ with field strength $\hat F_1=d\hat C_0$, dilaton $\hat\phi$; the RR and NSNS three-form field strengths $\hat{F}_3$ and $\hat{H}_3$%
\footnote{We use hats to denote string frame fields, while un-hatted fields denote Einstein frame fields as used in section \ref{sec:10Dsetup}.}%
. The Bianchi identities of these fields are:
\bse
\begin{align}
 0 &= d\hat{F}_1,\\
 0 &= d\hat{F}_3 - \hat{H}_3\wedge \hat{F}_1,\\
 0 &= d\hat{H}_3,
\end{align}
\ese
of which the last two are automatically satisfied when we introduce the (real) two-form gauge potentials $\hat{C}_2$ and $\hat{B}_2$ as:
\bse
\begin{align}
 \hat F_3 &= d\hat{C}_2 - \hat H_3 \hat C_0,\\
 \hat H_3 &= d\hat{B}_2.
\end{align}
\ese
The map between the complex fields $B$, $C_2$ used in section \ref{sec:10Dsetup} and the real fields $\hat C_0$, $\hat\phi$, $\hat{C}_2$, $\hat{B}_2$ is:
\bse
\begin{align}
 B&=\frac{1+i\tau}{1-i\tau}, & \tau&=\hat C_0+ie^{-\hat\phi},\\
 C_2&=\hat B_2+i\hat C_2.
\end{align}
\ese
The string frame metric $\hat{g}_{MN}$ is related to the Einstein frame metric (used in section \ref{sec:10Dsetup}) as:
\be \hat{g}_{MN} = e^{\hat\phi/2}\, g_{MN}.\ee
The equations of motion for the real 10D fields in string frame (including the self-dual five-form $F_5$\footnote{Often a different normalization for $F_5$ is used, e.g. $\hat{F}_5 = 4 F_5$ in \cite{Jeong:2013jfc}.}) are then given by (with $\hat*$ denoting the string frame Hodge dual):
\bse
\begin{align}
	0=&\,d(e^{-2\hat\phi}\hat*\hat H_3)-\hat F_1\wedge\hat* \hat F_3-4\hat 
    F_3\wedge\hat*F_5,\label{2B:eom:string:1}\\
	0=&\,d\hat*\hat F_1+\hat H_3\wedge\hat*\hat F_3,\label{2B:eom:string:2}\\
	0=&\,d\hat*\hat F_3+4\hat H_3\wedge\hat*F_5,\label{2B:eom:string:3}\\
	0=&\,\hat R+4\hat\nabla^2\hat\phi-4(\partial\hat\phi)^2-\frac{1}{12}(\hat H_3)^2,\label{2B:eom:string:4}\\
	0=&\,\hat R_{MN}+2\hat\nabla_M\hat\nabla_N\hat\phi-\frac{1}{4}(\hat H_3)^2_{MN}\nn\\&-e^{2\hat\phi}\left[\frac{1}{2}(\hat F_1)^2_{MN}+\frac{1}{4}(\hat F_3)^2_{MN}+\frac{1}{6}(F_5)^2_{MN}-\frac{1}{4}\hat g_{MN}(\hat F_1^2+\frac{1}{6}\hat F_3^2)\right],\label{2B:eom:string:5}\\
	0=&\,dF_5-\frac14\hat H_3\wedge\hat F_3,\label{2B:eom:string:6}
\end{align}
\label{2B:eoms:string}\ese
with the obvious notation of e.g.:
\be (\hat H_3)^2_{MN} = (\hat H_3)\ind{_M^{KL}} (\hat H_3)\ind{_{NKL}}.\ee 

\section{Checking the 10D EOMs}\label{app:EOMchecks}

We have verified that the 10D equations of motion (\ref{2B:eoms}) and (\ref{2B:eom:Einstein}) on the general reduction ansatz (\ref{general:ansatz:1})-(\ref{general:ansatz:G5}) are equivalent to the 6D equations of motion (\ref{F(4):eoms}) and (\ref{F(4):eom:5}). Here we give some details of this tedious but straightforward calculation.

The basic strategy is to write the forms (and their wedge products, Hodge duals, and exterior derivatives) as much as possible in terms of the derived quantities $d\calG, d(\calG/\mathcal{Y})$ and their 2D Hodge duals on the Riemann surface $*_2 d\calG, *_2d(\mathcal G/\mathcal{Y})$. Then, we can use some simple identities that are valid on the Riemann surface, such as:
\bse
\begin{align}
	d *_2 d\calG &= 2i\kappa^2 dz\wedge d\zb,\\
	d\calG\wedge *_2d\calG &= -2 i \kappa^2 \frac{\calG}{\mathcal{Y}} dz\wedge d\zb,\\
	\left(\frac{\calG}{\mathcal{Y}}\right)d *_2 d\left(\frac{\calG}{\mathcal{Y}}\right) &=3d\mathcal G\wedge*_2d\mathcal G+ 4d\calG\wedge*_2d\left(\frac{\calG}{\mathcal{Y}}\right) + 2d\left(\frac{\calG}{\mathcal{Y}}\right)\wedge*_2d\left(\frac{\calG}{\mathcal{Y}}\right).
\end{align}
\ese
One also needs to utilize a number of identities that the coordinates $\mu^i$ on the $S^2$ satisfy together with the gauge fields $\tilde F^i$, such as:
\bse
\begin{align}
 d \mathrm{vol}(\tilde S^2) &=d(\mu^i \tilde F^i)=\tilde F^i\wedge D\mu^i,\\
 D\mu^i\wedge \mathrm{vol}(\tilde S^2) &= 0,\\
 \epsilon_{ijk} D\mu^j\wedge D\mu^k &= 2\mu^i \mathrm{vol}(\tilde S^2),\\
 d(\tilde F^i\wedge *_2 D\mu^i) &= 2\mu^i \tilde F^i\wedge \mathrm{vol}(\tilde S^2) + \tilde F^i\wedge \tilde F^i - \mu^i\tilde F^i\wedge \mu^j \tilde F^j.
\end{align}
\ese

The IIB five-form and metric are given directly in (\ref{general:ansatz:4}), (\ref{general:ansatz:G5}), (\ref{general:ansatz:1}) and (\ref{eq:fXs}).  For the remaining fields $P$, $Q$ and $G_3$, we first note that the IIB equations of motion are invariant under the $U(1)$ transformation:
\be
    P\to e^{2i\theta}P,\qquad Q\to Q+d\theta,\qquad G_3\to e^{i\theta}G_3,
\label{U(1)}.
\ee
Inserting the expressions (\ref{general:ansatz:2}) and (\ref{general:ansatz:3}) for $B$ and $C_2$ into the expressions (\ref{eq:10D:PG3def}), and furthermore performing a $U(1)$ transformation with phase
\begin{equation}
	e^{i\theta}=\frac{(\mathcal A_+-\bar{\mathcal A}_-)+\hat CX^2/\sqrt{D}}{|(\mathcal A_+-\bar{\mathcal A}_-)+\hat CX^2/\sqrt{D}|},
\end{equation}
then gives
\bse
\begin{align}
	P=&-\fft2{3\mathcal Y\mathcal D}\fft{dX}X+\fft{\mathcal Y\sqrt{\mathcal D}}{2\mathcal GX^2}\left[i*_2\mathcal J-\fft{X^2}{\sqrt{\mathcal D}}\mathcal K\right],\label{2B:P1:U(1)}\\
	Q=&\fft{\mathcal Y\sqrt{\mathcal D}}{2\mathcal GX^2}*_2\mathcal L,\label{2B:Q1:U(1)}\\
	G_3=&\left(\fft{\mathcal Y\sqrt{\mathcal D}}{\mathcal GX^2}\right)^{\fft{1}{2}}\left[\left(-\fft{8ic_6\mathcal GX^3}{9\mathcal Y\mathcal D^2}dX+\fft{c_6}{3}\left(i\mathcal M+\fft{X^2}{\sqrt{\mathcal D}}*_2\mathcal N\right)\right)\wedge\mathrm{vol}(\tilde S^2)\right.\nn\\
	&\kern6em+2c_6\fft{\mathcal GX^2}{\mathcal Y\sqrt{\mathcal D}}\tilde F_3-\fft{c_6}{\sqrt2}\tilde F_2\wedge\left(i*_2d\mathcal G+\fft{X^2}{\sqrt{\mathcal D}}d\mathcal G\right)\nn\\
	&\kern6em\left.+\fft{c_6}3\left(\fft{2i}3\fft{\mathcal G}{\mathcal Y\mathcal D}\tilde F^i\wedge D\mu^i+\mu^i\tilde F^i\wedge\left(id\mathcal G-\fft{X^2}{\sqrt{\mathcal D}}*_2d\mathcal G\right)\right)\right],\label{eq:G3:U(1)}
\end{align}\label{2B:U(1)}\ese
where we have defined the shorthand one-forms $\mathcal{J}$, $\mathcal{K}$, $\mathcal L$, $\mathcal M$, $\mathcal{N}$ as:
\begin{subequations}
	\begin{align}
	\mathcal J&=\left(1-\fft{1}{\mathcal Y\mathcal D}\right)d\mathcal G+\left(1-\fft{2}{3\mathcal Y\mathcal D}\right)d\left(\fft{\mathcal G}{\mathcal Y}\right),\\
	\mathcal K&=\left(1+\fft{1}{3\mathcal Y^2\mathcal D}\right)d\mathcal G+\left(1-\fft{1}{3\mathcal Y\mathcal D}\right)d\left(\fft{\mathcal G}{\mathcal Y}\right),\\
	\mathcal L&=\left(2-\fft{1}{\mathcal Y\mathcal D}\right)d\mathcal G+\left(1-\fft{2}{3\mathcal Y\mathcal D}\right)d\left(\fft{\mathcal G}{\mathcal Y}\right),\\
	\mathcal M&=\left(\fft{4}{9\mathcal Y^2\mathcal D^2}+\fft{1}{3\mathcal D}-1\right)d\mathcal G+\fft{2}{3\mathcal D}\left(1-\fft{2}{3\mathcal Y\mathcal D}\right)d\left(\fft{\mathcal G}{\mathcal Y}\right),\\
	\mathcal N&=\left(1+\fft{1}{\mathcal D}\right)d\mathcal G+\fft{2}{3\mathcal D}d\left(\fft{\mathcal G}{\mathcal Y}\right).
	\end{align}
\end{subequations}

Now it is just a matter of using the above expressions to check the IIB equations of motion.  For example, using the expressions (\ref{2B:U(1)}) in the axi-dilaton equation of motion, (\ref{2B:eom:P}), gives
\begin{align}
	0=&-\fft{8ic_6^4\kappa^2\mathcal G}{9\mathcal Y\mathcal D}\left[d(X^{-1}*_6dX)-\fft{1}{4}X^4\left(\tilde F_3\wedge*_6\tilde F_3\right)-\fft{1}{8}X^{-2}\left(\tilde F_2\wedge*_6\tilde F_2+\fft{2}{9}\tilde F^i\wedge*_6\tilde F^i\right)\right.\nn\\&\left.\kern5em~-\fft{9}{2}\left(\fft{1}{6}X^{-6}-\fft{2}{3}X^{-2}+\fft{1}{2}X^2\right)(*_6\mathbbm{1})\right]\wedge\mathrm{vol}(\tilde S^2)\wedge dz\wedge d\zb,\label{P1:reduced}
\end{align}
which is equivalent to the 6D scalar equation of motion (\ref{F(4):eom:4}). The other equations of motion for the form-fields (\ref{2B:eom:G3}) and (\ref{2B:eom:F5}) proceed similarly (although they are more involved).

Finally, the Einstein equations are perhaps the most tedious of all to check. In the end, the $(\mu\nu)$ components of the 10D Einstein equations reduce to the 6D Einstein equations (\ref{F(4):eom:5}) and scalar equation (\ref{F(4):eom:4}) of motion. The $(z\zb)$ component of the Einstein equations reduces to the scalar equation (\ref{F(4):eom:4}), as do the components with both legs on the $S^2$. Finally, the components with one leg in the 6D manifold and one leg on the $S^2$ reduces to the equation of motion (\ref{F(4):eom:3}) for the $SU(2)$ gauge field. (All other components of the Einstein equations reduce to identities.)

\bibliographystyle{JHEP}
\bibliography{F4refs}

\providecommand{\href}[2]{#2}\begingroup\raggedright\begin{thebibliography}{10}

\bibitem{Nahm:1977tg}
W.~Nahm, \emph{{Supersymmetries and their Representations}},
  \href{https://doi.org/10.1016/0550-3213(78)90218-3}{\emph{Nucl. Phys.}
  {\bfseries B135} (1978) 149}.

\bibitem{Minwalla:1997ka}
S.~Minwalla, \emph{{Restrictions imposed by superconformal invariance on
  quantum field theories}},
  \href{https://doi.org/10.4310/ATMP.1998.v2.n4.a4}{\emph{Adv. Theor. Math.
  Phys.} {\bfseries 2} (1998) 783--851},
  [\href{https://arxiv.org/abs/hep-th/9712074}{{\ttfamily hep-th/9712074}}].

\bibitem{Romans:1985tw}
L.~J. Romans, \emph{{The F(4) Gauged Supergravity in Six-dimensions}},
  \href{https://doi.org/10.1016/0550-3213(86)90517-1}{\emph{Nucl. Phys.}
  {\bfseries B269} (1986) 691}.

\bibitem{Figueroa-OFarrill:2002ecq}
J.~M. Figueroa-O'Farrill and G.~Papadopoulos, \emph{{Maximally supersymmetric
  solutions of ten-dimensional and eleven-dimensional supergravities}},
  \href{https://doi.org/10.1088/1126-6708/2003/03/048}{\emph{JHEP} {\bfseries
  03} (2003) 048}, [\href{https://arxiv.org/abs/hep-th/0211089}{{\ttfamily
  hep-th/0211089}}].

\bibitem{Seiberg:1996bd}
N.~Seiberg, \emph{{Five-dimensional SUSY field theories, nontrivial fixed
  points and string dynamics}},
  \href{https://doi.org/10.1016/S0370-2693(96)01215-4}{\emph{Phys. Lett.}
  {\bfseries B388} (1996) 753--760},
  [\href{https://arxiv.org/abs/hep-th/9608111}{{\ttfamily hep-th/9608111}}].

\bibitem{Brandhuber:1999np}
A.~Brandhuber and Y.~Oz, \emph{{The D4-D8 brane system and five-dimensional
  fixed points}},
  \href{https://doi.org/10.1016/S0370-2693(99)00763-7}{\emph{Phys. Lett.}
  {\bfseries B460} (1999) 307--312},
  [\href{https://arxiv.org/abs/hep-th/9905148}{{\ttfamily hep-th/9905148}}].

\bibitem{Bergman:2012kr}
O.~Bergman and D.~Rodriguez-Gomez, \emph{{5d quivers and their $AdS_6$ duals}},
  \href{https://doi.org/10.1007/JHEP07(2012)171}{\emph{JHEP} {\bfseries 07}
  (2012) 171}, [\href{https://arxiv.org/abs/1206.3503}{{\ttfamily 1206.3503}}].

\bibitem{Passias:2012vp}
A.~Passias, \emph{{A note on supersymmetric AdS$_6$ solutions of massive type
  IIA supergravity}},
  \href{https://doi.org/10.1007/JHEP01(2013)113}{\emph{JHEP} {\bfseries 01}
  (2013) 113}, [\href{https://arxiv.org/abs/1209.3267}{{\ttfamily 1209.3267}}].

\bibitem{Aharony:1997ju}
O.~Aharony and A.~Hanany, \emph{{Branes, superpotentials and superconformal
  fixed points}},
  \href{https://doi.org/10.1016/S0550-3213(97)00472-0}{\emph{Nucl. Phys.}
  {\bfseries B504} (1997) 239--271},
  [\href{https://arxiv.org/abs/hep-th/9704170}{{\ttfamily hep-th/9704170}}].

\bibitem{Aharony:1997bh}
O.~Aharony, A.~Hanany and B.~Kol, \emph{{Webs of $(p,q)$ five-branes,
  five-dimensional field theories and grid diagrams}},
  \href{https://doi.org/10.1088/1126-6708/1998/01/002}{\emph{JHEP} {\bfseries
  01} (1998) 002}, [\href{https://arxiv.org/abs/hep-th/9710116}{{\ttfamily
  hep-th/9710116}}].

\bibitem{DeWolfe:1999hj}
O.~DeWolfe, A.~Hanany, A.~Iqbal and E.~Katz, \emph{{Five-branes, seven-branes
  and five-dimensional $E_n$ field theories}},
  \href{https://doi.org/10.1088/1126-6708/1999/03/006}{\emph{JHEP} {\bfseries
  03} (1999) 006}, [\href{https://arxiv.org/abs/hep-th/9902179}{{\ttfamily
  hep-th/9902179}}].

\bibitem{Apruzzi:2014qva}
F.~Apruzzi, M.~Fazzi, A.~Passias, D.~Rosa and A.~Tomasiello, \emph{{AdS$_{6}$
  solutions of type II supergravity}},
  \href{https://doi.org/10.1007/JHEP11(2014)099,
  10.1007/JHEP05(2015)012}{\emph{JHEP} {\bfseries 11} (2014) 099},
  [\href{https://arxiv.org/abs/1406.0852}{{\ttfamily 1406.0852}}].

\bibitem{Kim:2015hya}
H.~Kim, N.~Kim and M.~Suh, \emph{{Supersymmetric AdS$_6$ Solutions of Type IIB
  Supergravity}},
  \href{https://doi.org/10.1140/epjc/s10052-015-3705-1}{\emph{Eur. Phys. J.}
  {\bfseries C75} (2015) 484},
  [\href{https://arxiv.org/abs/1506.05480}{{\ttfamily 1506.05480}}].

\bibitem{DHoker:2016ujz}
E.~D'Hoker, M.~Gutperle, A.~Karch and C.~F. Uhlemann, \emph{{Warped
  $AdS_6\times S^2$ in Type IIB supergravity I: Local solutions}},
  \href{https://doi.org/10.1007/JHEP08(2016)046}{\emph{JHEP} {\bfseries 08}
  (2016) 046}, [\href{https://arxiv.org/abs/1606.01254}{{\ttfamily
  1606.01254}}].

\bibitem{DHoker:2017mds}
E.~D'Hoker, M.~Gutperle and C.~F. Uhlemann, \emph{{Warped $AdS_6\times S^2$ in
  Type IIB supergravity II: Global solutions and five-brane webs}},
  \href{https://doi.org/10.1007/JHEP05(2017)131}{\emph{JHEP} {\bfseries 05}
  (2017) 131}, [\href{https://arxiv.org/abs/1703.08186}{{\ttfamily
  1703.08186}}].

\bibitem{DHoker:2017zwj}
E.~D'Hoker, M.~Gutperle and C.~F. Uhlemann, \emph{{Warped $AdS_6\times S^2$ in
  Type IIB supergravity III: Global solutions with seven-branes}},
  \href{https://doi.org/10.1007/JHEP11(2017)200}{\emph{JHEP} {\bfseries 11}
  (2017) 200}, [\href{https://arxiv.org/abs/1706.00433}{{\ttfamily
  1706.00433}}].

\bibitem{Cvetic:1999un}
M.~Cvetic, H.~Lu and C.~N. Pope, \emph{{Gauged six-dimensional supergravity
  from massive type IIA}},
  \href{https://doi.org/10.1103/PhysRevLett.83.5226}{\emph{Phys. Rev. Lett.}
  {\bfseries 83} (1999) 5226--5229},
  [\href{https://arxiv.org/abs/hep-th/9906221}{{\ttfamily hep-th/9906221}}].

\bibitem{Jeong:2013jfc}
J.~Jeong, O.~Kelekci and E.~O~Colgain, \emph{{An alternative IIB embedding of
  F(4) gauged supergravity}},
  \href{https://doi.org/10.1007/JHEP05(2013)079}{\emph{JHEP} {\bfseries 05}
  (2013) 079}, [\href{https://arxiv.org/abs/1302.2105}{{\ttfamily 1302.2105}}].

\bibitem{Lozano:2012au}
Y.~Lozano, E.~\'O~Colg\'ain, D.~Rodr\'iguez-G\'omez and K.~Sfetsos,
  \emph{{Supersymmetric $AdS_6$ via T Duality}},
  \href{https://doi.org/10.1103/PhysRevLett.110.231601}{\emph{Phys. Rev. Lett.}
  {\bfseries 110} (2013) 231601},
  [\href{https://arxiv.org/abs/1212.1043}{{\ttfamily 1212.1043}}].

\bibitem{Duff:1985jd}
M.~J. Duff and C.~N. Pope, \emph{Consistent truncations in {K}aluza-{K}lein
  theories}, \href{https://doi.org/10.1016/0550-3213(85)90140-3}{\emph{Nucl.
  Phys.} {\bfseries B255} (1985) 355--364}.

\bibitem{Gauntlett:2007ma}
J.~P. Gauntlett and O.~Varela, \emph{{Consistent Kaluza-Klein reductions for
  general supersymmetric AdS solutions}},
  \href{https://doi.org/10.1103/PhysRevD.76.126007}{\emph{Phys. Rev.}
  {\bfseries D76} (2007) 126007},
  [\href{https://arxiv.org/abs/0707.2315}{{\ttfamily 0707.2315}}].

\bibitem{Malek:2017njj}
E.~Malek, \emph{{Half-Maximal Supersymmetry from Exceptional Field Theory}},
  \href{https://doi.org/10.1002/prop.201700061}{\emph{Fortsch. Phys.}
  {\bfseries 65} (2017) 1700061},
  [\href{https://arxiv.org/abs/1707.00714}{{\ttfamily 1707.00714}}].

\bibitem{Malek:2018zcz}
E.~Malek, H.~Samtleben and V.~Vall~Camell, \emph{{Supersymmetric AdS$_{7}$ and
  AdS$_6$ vacua and their minimal consistent truncations from exceptional field
  theory}},  \href{https://arxiv.org/abs/1808.05597}{{\ttfamily 1808.05597}}.

\bibitem{Schwarz:1983qr}
J.~H. Schwarz, \emph{{Covariant Field Equations of Chiral $N=2$ $D=10$
  Supergravity}},
  \href{https://doi.org/10.1016/0550-3213(83)90192-X}{\emph{Nucl. Phys.}
  {\bfseries B226} (1983) 269}.

\bibitem{Howe:1983sra}
P.~S. Howe and P.~C. West, \emph{{The Complete $N=2$, $D=10$ Supergravity}},
  \href{https://doi.org/10.1016/0550-3213(84)90472-3}{\emph{Nucl. Phys.}
  {\bfseries B238} (1984) 181--220}.

\bibitem{Apruzzi:2018cvq}
F.~Apruzzi, J.~C. Geipel, A.~Legramandi, N.~T. Macpherson and M.~Zagermann,
  \emph{{Minkowski$_4\times S^2$ solutions of IIB supergravity}},
  \href{https://doi.org/10.1002/prop.201800006}{\emph{Fortsch. Phys.}
  {\bfseries 66} (2018) 1800006},
  [\href{https://arxiv.org/abs/1801.00800}{{\ttfamily 1801.00800}}].

\bibitem{DHoker:2016ysh}
E.~D'Hoker, M.~Gutperle and C.~F. Uhlemann, \emph{{Holographic duals for
  five-dimensional superconformal quantum field theories}},
  \href{https://doi.org/10.1103/PhysRevLett.118.101601}{\emph{Phys. Rev. Lett.}
  {\bfseries 118} (2017) 101601},
  [\href{https://arxiv.org/abs/1611.09411}{{\ttfamily 1611.09411}}].

\bibitem{Gutperle:2018vdd}
M.~Gutperle, A.~Trivella and C.~F. Uhlemann, \emph{{Type IIB 7-branes in warped
  AdS$_{6}$: partition functions, brane webs and probe limit}},
  \href{https://doi.org/10.1007/JHEP04(2018)135}{\emph{JHEP} {\bfseries 04}
  (2018) 135}, [\href{https://arxiv.org/abs/1802.07274}{{\ttfamily
  1802.07274}}].

\bibitem{Bergman:2018hin}
O.~Bergman, D.~Rodr\'iguez-G\'omez and C.~F. Uhlemann, \emph{{Testing
  $AdS_6/CFT_5$ in Type IIB with stringy operators}},
  \href{https://arxiv.org/abs/1806.07898}{{\ttfamily 1806.07898}}.

\bibitem{Gutperle:2017tjo}
M.~Gutperle, C.~Marasinou, A.~Trivella and C.~F. Uhlemann, \emph{{Entanglement
  entropy vs. free energy in IIB supergravity duals for 5d SCFTs}},
  \href{https://doi.org/10.1007/JHEP09(2017)125}{\emph{JHEP} {\bfseries 09}
  (2017) 125}, [\href{https://arxiv.org/abs/1705.01561}{{\ttfamily
  1705.01561}}].

\bibitem{Gutperle:2018wuk}
M.~Gutperle, C.~F. Uhlemann and O.~Varela, \emph{{Massive spin 2 excitations in
  $AdS_6\times S^2$ warped spacetimes}},
  \href{https://doi.org/10.1007/JHEP07(2018)091}{\emph{JHEP} {\bfseries 07}
  (2018) 091}, [\href{https://arxiv.org/abs/1805.11914}{{\ttfamily
  1805.11914}}].

\bibitem{Fluder:2018chf}
M.~Fluder and C.~F. Uhlemann, \emph{{Precision test of AdS$_6$/CFT$_5$ in Type
  IIB}},  \href{https://arxiv.org/abs/1806.08374}{{\ttfamily 1806.08374}}.

\bibitem{Ferrara:1998gv}
S.~Ferrara, A.~Kehagias, H.~Partouche and A.~Zaffaroni, \emph{{AdS$_6$
  interpretation of 5-D superconformal field theories}},
  \href{https://doi.org/10.1016/S0370-2693(98)00560-7}{\emph{Phys. Lett.}
  {\bfseries B431} (1998) 57--62},
  [\href{https://arxiv.org/abs/hep-th/9804006}{{\ttfamily hep-th/9804006}}].

\bibitem{DAuria:2000afl}
R.~D'Auria, S.~Ferrara and S.~Vaula, \emph{{Matter coupled F(4) supergravity
  and the AdS$_6$/CFT$_5$ correspondence}},
  \href{https://doi.org/10.1088/1126-6708/2000/10/013}{\emph{JHEP} {\bfseries
  10} (2000) 013}, [\href{https://arxiv.org/abs/hep-th/0006107}{{\ttfamily
  hep-th/0006107}}].

\bibitem{Karndumri:2012vh}
P.~Karndumri, \emph{{Holographic RG flows in six dimensional F(4) gauged
  supergravity}}, \href{https://doi.org/10.1007/JHEP01(2013)134,
  10.1007/JHEP06(2015)165}{\emph{JHEP} {\bfseries 01} (2013) 134},
  [\href{https://arxiv.org/abs/1210.8064}{{\ttfamily 1210.8064}}].

\bibitem{Karndumri:2014lba}
P.~Karndumri, \emph{{Gravity duals of 5D $N=2$ SYM theory from $F(4)$ gauged
  supergravity}}, \href{https://doi.org/10.1103/PhysRevD.90.086009}{\emph{Phys.
  Rev.} {\bfseries D90} (2014) 086009},
  [\href{https://arxiv.org/abs/1403.1150}{{\ttfamily 1403.1150}}].

\bibitem{Alday:2014rxa}
L.~F. Alday, M.~Fluder, P.~Richmond and J.~Sparks, \emph{{Gravity Dual of
  Supersymmetric Gauge Theories on a Squashed Five-Sphere}},
  \href{https://doi.org/10.1103/PhysRevLett.113.141601}{\emph{Phys. Rev. Lett.}
  {\bfseries 113} (2014) 141601},
  [\href{https://arxiv.org/abs/1404.1925}{{\ttfamily 1404.1925}}].

\bibitem{Alday:2014bta}
L.~F. Alday, M.~Fluder, C.~M. Gregory, P.~Richmond and J.~Sparks,
  \emph{{Supersymmetric gauge theories on squashed five-spheres and their
  gravity duals}}, \href{https://doi.org/10.1007/JHEP09(2014)067}{\emph{JHEP}
  {\bfseries 09} (2014) 067},
  [\href{https://arxiv.org/abs/1405.7194}{{\ttfamily 1405.7194}}].

\bibitem{Alday:2014fsa}
L.~F. Alday, P.~Richmond and J.~Sparks, \emph{{The holographic supersymmetric
  Renyi entropy in five dimensions}},
  \href{https://doi.org/10.1007/JHEP02(2015)102}{\emph{JHEP} {\bfseries 02}
  (2015) 102}, [\href{https://arxiv.org/abs/1410.0899}{{\ttfamily 1410.0899}}].

\bibitem{Hama:2014iea}
N.~Hama, T.~Nishioka and T.~Ugajin, \emph{{Supersymmetric R\'enyi entropy in
  five dimensions}}, \href{https://doi.org/10.1007/JHEP12(2014)048}{\emph{JHEP}
  {\bfseries 12} (2014) 048},
  [\href{https://arxiv.org/abs/1410.2206}{{\ttfamily 1410.2206}}].

\bibitem{Alday:2015jsa}
L.~F. Alday, M.~Fluder, C.~M. Gregory, P.~Richmond and J.~Sparks,
  \emph{{Supersymmetric solutions to Euclidean Romans supergravity}},
  \href{https://doi.org/10.1007/JHEP02(2016)100}{\emph{JHEP} {\bfseries 02}
  (2016) 100}, [\href{https://arxiv.org/abs/1505.04641}{{\ttfamily
  1505.04641}}].

\bibitem{Gutperle:2017nwo}
M.~Gutperle, J.~Kaidi and H.~Raj, \emph{{Janus solutions in six-dimensional
  gauged supergravity}},
  \href{https://doi.org/10.1007/JHEP12(2017)018}{\emph{JHEP} {\bfseries 12}
  (2017) 018}, [\href{https://arxiv.org/abs/1709.09204}{{\ttfamily
  1709.09204}}].

\bibitem{Chang:2017mxc}
C.-M. Chang, M.~Fluder, Y.-H. Lin and Y.~Wang, \emph{{Romans Supergravity from
  Five-Dimensional Holograms}},
  \href{https://doi.org/10.1007/JHEP05(2018)039}{\emph{JHEP} {\bfseries 05}
  (2018) 039}, [\href{https://arxiv.org/abs/1712.10313}{{\ttfamily
  1712.10313}}].

\bibitem{Gutperle:2018axv}
M.~Gutperle, J.~Kaidi and H.~Raj, \emph{{Mass deformations of 5d SCFTs via
  holography}}, \href{https://doi.org/10.1007/JHEP02(2018)165}{\emph{JHEP}
  {\bfseries 02} (2018) 165},
  [\href{https://arxiv.org/abs/1801.00730}{{\ttfamily 1801.00730}}].

\bibitem{Duff:1984hn}
M.~J. Duff, B.~E.~W. Nilsson, C.~N. Pope and N.~P. Warner, \emph{{On the
  Consistency of the {Kaluza-Klein} Ansatz}},
  \href{https://doi.org/10.1016/0370-2693(84)91558-2}{\emph{Phys. Lett.}
  {\bfseries 149B} (1984) 90--94}.

\bibitem{Hoxha:2000jf}
P.~Hoxha, R.~R. Martinez-Acosta and C.~N. Pope, \emph{{Kaluza-Klein
  consistency, Killing vectors, and Kahler spaces}},
  \href{https://doi.org/10.1088/0264-9381/17/20/305}{\emph{Class. Quant. Grav.}
  {\bfseries 17} (2000) 4207--4240},
  [\href{https://arxiv.org/abs/hep-th/0005172}{{\ttfamily hep-th/0005172}}].

\end{thebibliography}\endgroup

\end{document}